\RequirePackage[2020-02-02]{latexrelease}
\documentclass
[aps,preprint,showkeys,a4paper,tightenlines,superscriptaddress]{revtex4-2}%
\usepackage{eurosym}
\usepackage{amsfonts}
\usepackage{amsmath}
\usepackage{amssymb}
\usepackage{graphicx}
\usepackage{placeins}
\usepackage{textcomp}
\usepackage{gensymb}
\usepackage{subfig}
\usepackage{caption}
\usepackage[dvipsnames]{xcolor}%
\providecommand{\U}[1]{\protect\rule{.1in}{.1in}}
\providecommand{\U}[1]{\protect\rule{.1in}{.1in}}

\makeatletter

\renewcommand*{\p@subsection}{}

\renewcommand*{\p@subsubsection}{}
\makeatother

\begin{document}
\title{Modelling the non-steady peeling of viscoelastic tapes}
\author{M. Ceglie}
\affiliation{Department of Mechanics, Mathematics and Management, Politecnico of Bari, V.le
Japigia, 182, 70126, Bari, Italy}
\author{N. Menga}
\email{Corresponding author: nicola.menga@poliba.it}
\affiliation{Department of Mechanics, Mathematics and Management, Politecnico of Bari, V.le
Japigia, 182, 70126, Bari, Italy}
\author{G. Carbone}
\affiliation{Department of Mechanics, Mathematics and Management, Politecnico of Bari, V.le
Japigia, 182, 70126, Bari, Italy}
\affiliation{Imperial College London, Department of Mechanical Engineering, Exhibition
Road, London SW7 2AZ}
\affiliation{CNR - Institute for Photonics and Nanotechnologies U.O.S. Bari, Physics
Department \textquotedblright M. Merlin\textquotedblright, via Amendola 173,
70126 Bari, Italy}
\keywords{viscoelasticity, adhesion, peeling, bioinspired, detachment}
\begin{abstract}
We present a model to study the non-steady V-shaped peeling of a viscoelastic thin tape adhering to a rigid flat substrate. Geometry evolution and viscoelastic creep in the tape are the main features involved in the process, which allows to derive specific governing equations in the framework of energy balance. Finally, these are numerically integrated following an iterative scheme to calculate the process evolution assuming different controlling conditions (peeling front velocity, peeling force, tape tip velocity). Results show that the peeling behavior is strongly affected by viscoelasticity. Specifically, for a given applied force, the peeling can either be prevented, start and stop after some while, or endlessly propagate, depending on the original undeformed tape geometry. Viscoelasticity also entails that the interface toughness strongly increases when the tape tip is fast pulled, which agrees to recent experimental observations on tougher adhesion of natural systems under impact loads, such as see waves and wind gusts.

\end{abstract}
\maketitle

\begin{tabular}
[c]{|p{0.15\textwidth}p{0.75\textwidth}|}\hline
\multicolumn{2}{|l|}{\textbf{Nomenclature}}\\[0.5ex]
$A_{t}=wd$ & Tape cross section\\
$d$ & Tape thickness\\
$E_{0}$, $E_{\infty}$ & Low and high frequency viscoelastic moduli\\
$\mathcal{J}$ & Viscoelastic creep function\\
$L_{i}$ & initial undeformed length of the non-adhering tape\\
$N$ & Number of detached elements\\
$P$ & Peeling force\\
$\mathcal{R}$ & Viscoelastic relaxation function\\
$s_{d}$ & Undeformed peeled tape length\\
$v_{c}$ & Peeling front velocity\\
$v_{P}$ & Tape tip pulling velocity\\
$v_{\gamma}$ & Reference peeling velocity for adhesion\\
$w$ & Tape width\\
$W_{in}, W_{P}, W_{ad}$ & Power of the internal stress, the peeling force, and
the adhesive bonds\\
$\gamma$ & Adhesion energy\\
$\gamma_{0}$ & Nominal adhesion energy for $v_{c}\ll v_{\gamma}$\\
$\Delta L$ & Total detached tape elongation\\
$\Delta t$ & Time step\\
$\varepsilon$ & Tape deformation\\
$\theta$ & Peeling angle\\
$\phi$ & Peeling angle at rest\\
$\kappa=E_{\infty}/E_{0}$ & Viscoelastic parameter\\
$\lambda, \lambda_{c}$ & Tape joint coordinate and peeling front location\\
$\sigma$ & Tape stress\\
$\sigma_{cr}$ & Critical stress value to trigger peeling propagation\\
$\tau$ & Creep time\\
$\tau_{r}$ & Relaxation time\\
$P_{0}, \theta_{0}$ & Critical force and angle for peeling initiation\\
$P_{S}, \theta_{S}$ & Long-term steady-state peeling force and angle\\\hline
\end{tabular}

\section{Introduction}

In modern science, the study of attachment and detachment mechanisms is of practical importance for several applications, such as climbing ability in soft-robotics \cite{Creton2007,Cho2009}, deposition and removal of coating for specialized interfaces \cite{Min2008}, pick-and-place processes in manufacturing \cite{Meitl2006}, self-healing heterogeneous materials for construction \cite{Hwang2015} and wound dressing for medical industries \cite{Shafiee2021}. Among the others, when dealing with tapes and membranes, as well as fibrils and thin bristles, detachment through mechanical peeling has recently seen a growing interest, quickly becoming the main mechanism for systems such as electro-adhesive \cite{Cacucciolo2021} and shear-activated nano-structured \cite{Hawkes2017} grippers for objects manipulation made of compliant membranes, band-aids \cite{Kwak2011} and tunable skin patch \cite{Bae2013} to minimize the removal damage of biological tissues \cite{plaut2010,karwoski2004}, spray coatings \cite{Sexsmith1994} and transfer printing \cite{Wang2019,Feng2007} for flexible circuits fabrication (also adopting micro-fibrils adhesives\cite{Song2014}), and highly-stretchable structural adhesive tapes \cite{Sato2016}.

Since the first experimental study by Rivlin dealing with (almost) rigid tapes \cite{Rivlin1944}, a dramatic effort has been made to include the effect of tape deformability, with a specific focus on finite strain \cite{Eremeyev2015,Molinari2008}, prestrain \cite{Williams2004, Chen2009,Putignano2014}, bending stiffness \cite{Peng2015,Sauer2011}, tape plasticity \cite{Kim1988}, peel rate dependent adhesion \cite{Maugis1980,Kovalchick2014,Zhou2011}, and film-substrate interface conditions (e.g., frictional sliding and stick-slip) \cite{Begley2013,Dalbe2015,Camara2008,Maybhate2004}. It is the case, for instance, of energy-based Kendall's model \cite{Kendall1971} for elastic tapes peeling, which still represents the benchmark for a broad class of real systems, also including ISO standards for adhesive interfaces \cite{Pelfrene2015,Fafenrot2019}.

Although peeling itself is a local phenomenon involving a crack propagation at the interface between a layer and a substrate, the macroscopic detachment response is also affected by the global system properties. Those of main interest for similar tribological problems usually are the system geometry \cite{Menga2019,Menga2021,Müller2023} and the materials rheology \cite{Afferrante2019,Carbone2022}. Indeed, studying peeling geometries other than single peeling (in which case the peeling angle equals the force angle) has been recently urged by biomimetics in the attempt, for instance, to mimic \cite{Shahsavan2017} the superior locomotion performance of spiders, insects, and reptiles. This depends on the ability to quickly detach their hierarchical-structured toes by exploiting simultaneous peeling fronts propagation, ranging from the macro-scale (e.g., the leg) to the nano-scale (e.g., the toe spatula) \cite{Heepe2017}. Pugno and coworkers \cite{Pugno2011,Bosia2014} suggested that both the hierarchy and V-shape of the peeling geometry of such systems may play a key role in the overall toughness as multiple peeling fronts coexist, and the peeling angle varies during the detachment process. Later, Lepore et Al. \cite{Lepore2012} showed that the angles assumed by Tokay geckos at the two characteristic sizes of feet and toes are in excellent agreement with Pugno's multiple peeling theory predictions. Similarly, V-peeling geometry has been observed in spiders' webs anchors \cite{Cranford2012,Gu2016} and byssus threads networks of mussels \cite{Qin2013,Desmond2015}, both showing superior adhesive performance and the ability to withstand heavy winds and waves. Moreover, since peeling has also been successfully employed in characterizing the adhesive properties of materials and adhesives \cite{Bartlett2023}, as well as to assess the toughness of interfaces, specific tests (e.g., ASTM Loop Tack test) have been defined relying on the V-peeling geometry to reduce the possible effect of tape bending \cite{Gent1986}, compared to standard $90\degree$-$180\degree$ peel tests.

\begin{figure}[ptbh]
\begin{center}
\includegraphics[width=0.9\textwidth]{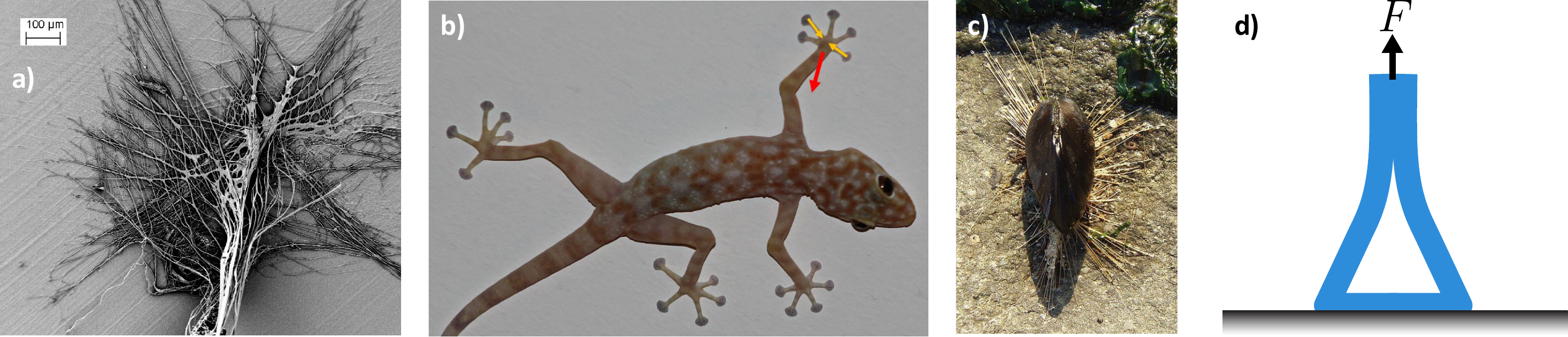}
\end{center}
\caption{Examples of V-peeling configurations in natural systems and practical
applications: (a) spider web anchors (from Ref. \cite{Menga2018}); (b) gecko
upside down climbing (adapted from Wikipedia); (c) mussel byssus threads (from
Wikipedia); (d) loop tack test schematic (from Ref. \cite{Bartlett2023}).}%
\label{image}%
\end{figure}

Nonetheless, existing models for V-peeling geometry only focus on elastic tapes \cite{Elder2020}, whereas biological systems and commercial tapes usually exhibit a certain degree of viscoelasticity.
Indeed, the effect of the materials viscoelasticity and, in turn, of energy dissipation during creep deformation, has been mostly addressed with reference to the single peeling configuration. Both physical \cite{Ceglie2022,Loukis1991,Chen2013} and phenomenological \cite{Maugis1980,Benyahia1997,Derail1997,Zhou2011,Peng2014} models have been developed, showing that tape viscoelasticity makes the peeling toughness increase with peel rate. More in detail, Ceglie et Al. \cite{Ceglie2022} have shown that viscoelasticity and local frictional sliding close to the peeling front may lead to unbounded peeling toughness at very low peeling angles, in agreement with existing experimental results \cite{Collino2014}.
Zhu et Al. \cite{Zhu2021a} focused on the tape visco-hyperelasticity effect on specific zero-degree peeling configuration, showing that for relatively thick tapes the peeling expected peeling force is less sensitive to bulk properties and interfacial adhesion (i.e., surface defects) compared to the case of linear rheology materials.

Viscoelasticity can also be localized in the substrate \cite{Afferrante2016} leading to an "ultra-tough" behavior achievable at specific peel rates. Similar results were also confirmed by Pierro et Al. \cite{Pierro2020} for a real viscoelastic material with a broader relaxation spectrum and by Zhu at Al. \cite{zhu2021b} also assuming rate dependent interface adhesion. The substrate rheology is crucial, for instance, when adhesive tapes are removed from human skin, in which case Renvoise at Al. \cite{renvoise2009} showed that at relatively large peeling velocity the multi-layer nature of human skin can also matter. Surprisingly, less has been done combining rheology effects and V-peeling geometry, although Menga et Al. \cite{Menga2018,Menga2020} have shown that, for purely elastic conditions, highly compliant substrates can drastically alter the peeling toughness in V-shaped systems due to the elastic interaction between adjacent peeling fronts.

In this study, we present a model for the V-peeling process of viscoelastic tapes backed onto rigid substrates, aiming at fostering the understanding of the mechanisms underlying the superior performance shown by insect toes, mussels attachment structures, and other biological systems relying on V-shaped peeling geometry in the presence of viscoelasticity, as well as to enhance the accuracy of loop tack test analysis to predict the adhesive performance of real interfaces. Since the process under investigation is non-steady, in Section 2 we set the appropriate theoretical energy-based framework and derive the governing equations for the peeling load, angle and front velocity, while the numerical procedure to integrate such equations and predict the process evolution over time is given in Appendix A and B. Results are presented in Section 3, focusing on three possible peeling procedures (constant peeling front velocity, constant peeling load, and constant velocity of the tape tip), each of which leads to qualitatively different results highlighting the interplay between V-shaped geometry and tape viscoelasticity.

\section{Formulation}
\label{Formulation}

\begin{figure}[ptbh]
\begin{center}
\includegraphics[width=1\textwidth]{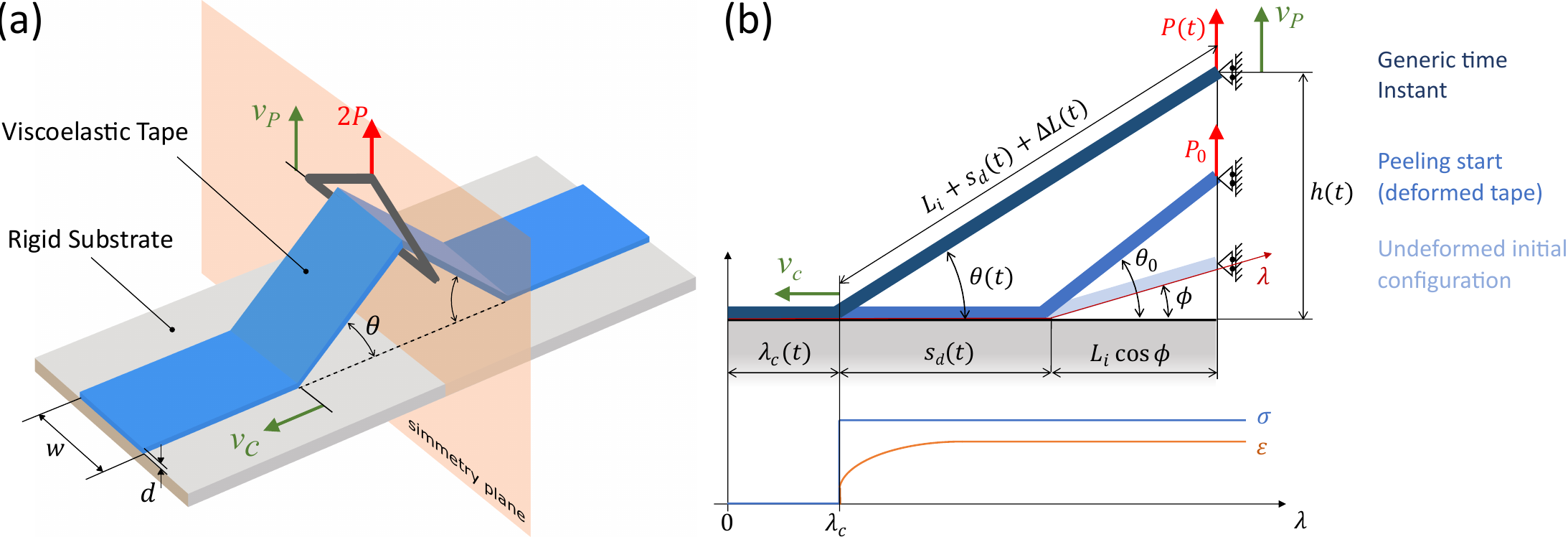}
\end{center}
\caption{(a) Double V-shaped peeling scheme of a viscoelastic tape adhering to a rigid substrate. $v_{c}$ is the peeling front propagation velocity, and
$v_{P}$ is the pulling velocity (i.e., the velocity of the tape tip). (b) By exploiting the system symmetry, the study only focuses on half of the tape. We show three different configurations: the undeformed tape, the tape at peeling propagation start (subscript $_0$), and a generic time instant with peeling force $P(t)$ and angle $\theta(t)$. In the bottom part, we also show qualitative diagrams of the stress $\sigma$ (blue) and deformation $\varepsilon$ (orange) along the tape coordinate $\lambda$.}
\label{fig1}%
\end{figure}

We consider the peeling configuration shown in Fig. \ref{fig1}a, where a thin viscoelastic tape of thickness $d$ and width $w$ adhering to a rigid substrate is pulled away by a normal force $2P$. Since two peeling fronts propagate in opposite directions (V-shaped double peeling), the whole process is symmetric with respect to the force direction and the study can be limited to half of the system, as shown in Fig. \ref{fig1}b. 

Before the peeling force $P$ being applied, the (undeformed) non-adhering tape length and angle are $L_{i}$ and $\phi$, respectively. Once the force is applied, at the time instant when the peeling starts to propagate, the tape angle is $\theta_0$ and the peeling force is $P_0$. According to Fig. \ref{fig1}b, at a generic time $t$, the peeling front coordinate and velocity are $\lambda_{c}(t)$ and $v_{c}\left(  t\right)  =-d\lambda_{c}/dt$, respectively, with $\lambda$ being the (undeformed) tape-fixed reference frame. Similarly, $s_{d}(  t)=\int_{0}^{t}v_c(  t)  dt$ is the detached tape length, and the peeling angle $\theta(t)$ is given as
\begin{equation}
\cos\theta=\frac{s_{d}+L_{i}\cos\phi}{L_{i}+s_{d}+\Delta L} ,
\label{geometric}
\end{equation}
where  $\Delta L(t) =\int_{\lambda_{c}}^{\lambda_{c}+s_{d}+L_{i}%
}\varepsilon(\lambda,t)d\lambda$\ is the elongation of the overall non-adhering tape, with $\varepsilon(\lambda,t)$ being the extensional deformation field in the tape. 

The instantaneous energy balance governing the peeling process is
\begin{equation}
W_{P}+W_{in}+W_{ad}=0{,}
\label{Energy Balance}
\end{equation}
where $W_{P}\left(  t\right)  $ is the work per unit time done by the peeling force $P\left(  t\right)  $, $W_{in}\left(  t\right)  $ is the work per unit time done by the internal stress field, and $W_{ad}\left(  t\right)  $ is the rate of the surface adhesion energy. In Eq. (\ref{Energy Balance}), minor energy contributions ascribable to acoustic emissions and heat transfer are neglected, as well as dynamic and inertial effects which might lead to stick-slip unstable delamination
\cite{Maugis1988,Cortet2007,Amouroux2001,Lake1981,Ciccotti2004}. Moreover, we assume a fully stuck adhesion between the tape and the rigid substrate in the adhering region, thus no friction energy dissipation occurs due to relative sliding, as instead considered in Refs. \cite{Ceglie2022,Begley2013}.

The term $W_{in}\left(  t\right)  $ is associated with both the rate of elastic energy stored in the detached tape and the viscoelastic energy loss occurring during the tape relaxation. Large deformations can be reasonably expected for soft polymeric tapes; however, both numerical \cite{Molinari2008} and experimental \cite{Collino2014} studies have clearly shown that real systems exhibiting strains as large as beyond $60\%$ can still be both qualitatively and quantitatively described in linear theory approximation, especially at relatively large peeling angles \cite{He2019}. Similar results are confirmed for visco-hyperelastic tapes \cite{Zhu2021a}, where qualitatively different behaviors are expected only beyond approximately $100\%$ strain value.
Moreover, we assume purely extensional stress $\sigma(\lambda,t)$ and deformation $\varepsilon(\lambda,t)$ fields in the tape, as experiments have shown that bending effects vanish for very thin tapes \cite{Gent1977} (i.e., the tape bending stiffness depends on $d^{3}$). Therefore, we have%
\begin{equation}
W_{in}=-A_{t}\int_{\lambda_{c}}^{\lambda_{c}+L_{i}+s_{d}}\sigma(\lambda
,t)\frac{\partial\varepsilon}{\partial t}(\lambda,t)d\lambda{,}
\label{win}
\end{equation}
with $\varepsilon(\lambda,t)=\sigma(\lambda,t)=0$ for $\lambda<\lambda_{c}$ (adhering tape) and $\sigma(\lambda,t)=\sigma(t)=P/\left(  A_{t}\sin
\theta\right)  $ for $\lambda>\lambda_{c}$ (detached tape). In the peeling section (i.e., for $\lambda=\lambda_{c}$), a step change of the stress occurs \cite{Ceglie2022}, so that
\begin{equation}
\sigma(\lambda,t)=\sigma(t)H[\lambda-\lambda_{c}(t)]{,}
\label{sigma}
\end{equation}
where $\mathcal{H}$ is the Heaviside step function. In the framework of linear viscoelasticity, the deformation field within the tape
is given by
\begin{equation}
\varepsilon(\lambda,t)=\int\nolimits_{-\infty}^{t}\mathcal{J}(t-t^{\prime
})\frac{\partial\sigma}{\partial t^{\prime}}(\lambda,t^{\prime})dt^{\prime}{,}
\label{constitutive}%
\end{equation}
where $\mathcal{J}$ is the viscoelastic creep function which, for a single
characteristic creep time $\tau$, is given by
\begin{equation}
\mathcal{J}(t)=\frac{1}{E_{0}}-\frac{e^{-t/\tau}}{E_{1}}{,}
\label{Creep function}%
\end{equation}
where $E_{1}^{-1}=E_{0}^{-1}-E_{\infty}^{-1}$, with $E_{0}$ and $E_{\infty}$
being the low and high frequency viscoelastic moduli, respectively.

The term $W_{P}\left(  t\right)  $ in Eq. (\ref{Energy Balance}) is given by
\begin{equation}
W_{P}=P\,v_{P}=\sigma A_{t}v_{P}\sin\theta {,} 
\label{wf}%
\end{equation}
where
\begin{equation}
v_{P}=\frac{dh}{dt}=v_{c}\,\tan\theta+\frac{s_{d}+L_{i}\cos\phi\,}{\cos
^{2}\theta}\dot{\theta} \label{velocity}%
\end{equation}
is the pulling velocity (see Fig. \ref{fig1}).

Finally, in Eq. (\ref{Energy Balance}), $W_{ad}\left(  t\right)  $ represents
the energy per unit time associated with the rupture of interfacial bonds
between the tape and the rigid substrate; being $\gamma$ the energy of
adhesion (also called Dupre's energy), we have%
\begin{equation}
W_{ad}=-v_{c}w\gamma{.}
\label{wad}
\end{equation}

The adhesion energy $\gamma$ might, in general, depend on the peeling
velocity, as reported by several experiments
\cite{Maugis1980,Peng2014,Feng2007,Zhou2011}. This is usually ascribed to
viscoelastic non-conservative (stiffening) effects in the tape close to the
peeling front, as recently predicted in Ref. \cite{Ceglie2022}. Here, we
precisely model the tape viscoelastic creep, thus the latter effect is
intrinsically accounted for. However, as pointed out by Marin \& Derail
\cite{Marin2006} with \textit{ad hoc} tests on inextensible tapes, velocity-dependent
power loss is also localized in the thin adhesive layer between the tape and
the substrate, with $\gamma$ given by the following power-law%
\begin{equation}
\gamma=\gamma_{0}\left[  1+\left(  \frac{v_{c}}{v_{\gamma}}\right)
^{n}\right]{,}
\label{adhesion energy}
\end{equation}
where $\gamma_{0}$ is the nominal adhesion energy for $v_{c}\ll v_{\gamma}$, with
$v_{\gamma}$ being a reference peeling velocity, and $n$ being a constant
which depends on the properties of the adhesive (typically in a range of
$0.3 - 0.7$) \cite{Marin2006,Feng2007,Peng2014}.

At any given time $t$, Eqs. (\ref{geometric},\ref{Energy Balance}) allow to
calculate the critical condition for peeling propagation.

To set the range of validity of the present model, we observe that the real deformation process occurring across the peeling front is continuous and cannot be formally represented by a step-change in the stress field. Nonetheless, physical arguments suggest that the length of the region  undergoing the stress increase from $0$ to $\sigma$ 
is of the same order of magnitude as the tape thickness $d$
\cite{Afferrante2016,Menga2020}, which results in a local excitation frequency $\omega\approx v_{c}/d$ and allows to identify
three different qualitative behaviors across the peeling section.  For $v_{c}\ll d/\tau$ (i.e., $\omega\ll1/\tau$), the tape behaves almost elastically, with elastic modulus approaching the
low-frequency modulus $E_{0}$. Since no viscoelastic dissipation occurs, this case is clearly out of the scope of this study, and the corresponding peeling behavior follows the elastic predictions given in Refs. \cite{Pugno2011,Afferrante2013}. 
For $v_{c}\approx d/\tau$ (i.e., $\omega\approx1/\tau$), the tape response strongly depends on the specific deformation process across the peeling front (small-scale energy dissipation cannot be neglected), and a local \textit{ad hoc} solid mechanics formulation is required to model the peeling.
Finally, the third case is the one of interest for the present model, as for $v_{c}\gg d/\tau$ (i.e., $\omega\gg1/\tau$)
the tape behavior is elastic across the peeling section with high frequency
elastic modulus $E_{\infty}$, and viscoelastic losses are localized in the
non-adhering tape (large-scale). Daily-life adhesive tapes are commonly very thin, with a corresponding threshold velocity usually being in the range of $d/\tau\approx10 - 100$ $\mu$m/s, which makes the third case of most relevant practical interest.

\subsection{Steady-state long-term propagation limit}

After the initial transient regime, the elastic V-peeling process asymptotically approaches a steady-state regime in the long-term limit \cite{Afferrante2013}. In the viscoelastic case, a similar behavior is expected for $t\gg\tau$ and $s_{d}\gg L_{i}$, which corresponds to complete viscoelastic relaxation along the tape. In this case, steady-state conditions occurs with $\theta(t)\approx\theta_{\mathrm{S}}$,  $\sigma(\lambda,t)\approx\sigma_{\mathrm{S}}=P_{\mathrm{S}}/\left(  A_{t}\sin\theta_{\mathrm{S}%
}\right)$, and $\varepsilon(\lambda,t)\approx\varepsilon_{\mathrm{S}}=\sigma_{\mathrm{S}%
}/E_{0}$, while the energy balance equation recovers the viscoelastic single peeling form \cite{Ceglie2022} as 
\begin{equation}
\frac{\sigma_{\mathrm{S}}^{2}}{2E_{\infty}}+\sigma_{\mathrm{S}}\left(
1-\cos\theta_{\mathrm{S}}\right)  -\frac{\gamma}{d}=0{,}
\label{steady state 1}%
\end{equation}
where $\gamma=\gamma(v_c)$ is given by Eq. (\ref{adhesion energy}). Moreover, since $s_{d}\gg L_{i}$, Eq. (\ref{geometric}) can be rewritten as 
\begin{equation}
\;\frac{1-\cos\theta_{\mathrm{S}}}{\cos\theta_{\mathrm{S}}}=\frac
{\sigma_{\mathrm{S}}}{E_{0}}{.}
\label{steady state 2}
\end{equation}
Furthermore, since $\dot{\theta}_{\mathrm{S}}\approx0$, Eq. (\ref{velocity}) gives
\begin{equation}
v_{P}=v_{c}\,\tan\theta_{\mathrm{S}}{.}
\label{steady state vel}
\end{equation}
Notably, under force-controlled conditions (i.e., given $P=P_{\mathrm{S}}$), Eqs. (\ref{steady state 1},\ref{steady state 2},\ref{adhesion energy}) allow to calculate the peeling front velocity $v_{c}$ and, through Eq. (\ref{steady state vel}), the pulling velocity $v_{P}$. On the contrary, under velocity-controlled conditions (i.e., given $v_{c}$ or $v_{P}$), the value of $P_{\mathrm{S}}$ can be calculated by Eqs. (\ref{steady state 1}%
,\ref{steady state 2},\ref{adhesion energy}).

\section{Results and Discussion}

In this section, we discuss the peeling behavior resulting from Eqs. (\ref{geometric},\ref{Energy Balance}), which can be numerically solved by following the procedure outlined in Appendix \ref{numerical implementation}. To simplify the analysis of the results, we refer to dimensionless quantities, i.e. $\tilde
{t}=t/\tau$,$\ \tilde{\gamma}=\gamma/E_{0}d$, $\tilde{v}_{c}=v_{c}\tau/d$, $\tilde{v}_{P}=v_{P}\tau/d$, $\tilde{P}=P/dwE_{0}=\sigma\sin\theta/E_{0}$. In our calculations, we consider a tape of thickness $d\approx100$ $\mu$m with initial non-adhering length $L_i=100 d$. The tape material is viscoelastic with low-frequency modulus $E_{0}=10$ MPa and creep time $\tau=1$ s.
Marin and Derail \cite{Marin2006} measured the peeling force $P$ as a function of the peeling velocity $v_c$ for real adhesives with inextensible aluminum backing, and Rivlin peeling theory \cite{Rivlin1944} allows to calculate the corresponding effect of $v_c$ on the adhesion energy $\gamma$. Since no viscoelastic relaxation occurs, the latter effect is only ascribable to
non-conservative phenomena localized in the very proximity of the peeling front. According to their results, we set $v_{\gamma}\approx10^{-3}$ m/s, $\gamma
_{0}\approx20$ J/m$^{2}$, and $n=0.5$ in Eq. \ref{adhesion energy}, whose corresponding dimensionless quantities are  
$\tilde{v}_{\gamma}=v_{\gamma
}\tau/d=10$ and $\tilde{\gamma}_{0}=\gamma_{0}/E_{0}d=0.02$.

Results are presented considering three different controlling parameters, corresponding to specific physical scenarios: (i) peeling propagation occurring at constant peeling front velocity $v_{c}$;
(ii) the case of a constant peeling force $P$ applied at the tape tip; and (iii) the case of the tape tip pulled at constant velocity $v_{P}$.

\subsection{Constant peeling front velocity}

We firstly consider the peeling process occurring at constant peeling front
velocity $v_{c}$. This case corresponds to time-varying values of both the peeling force $P$ and
pulling velocity $v_{P}$, thus resulting harder to be straightforwardly associated with common applications.
However, since $v_c$ also represents the length of undeformed tape that detaches the substrate per unit time and, once deformed, undergoes viscoelastic relaxation, fundamental insight on the interplay between peeling propagation and tape viscoelasticity. 

With reference to Fig. \ref{fig1}b, we assume that the peeling front propagation starts with velocity $v_c$ at time $t=0$ under the action of the critical force $P_{0}$, which is instantaneously applied. At this time, the tape deformation and angle undergo a step-change, varying from $\varepsilon=0$ and $\phi$ at time $t\rightarrow0^{-}$ to $\varepsilon=\sigma_{0}/E_{\infty
}$ and $\theta_{0}$ at time $t\rightarrow0^{+}$. 
As a consequence, no viscoelastic loss occurs in the non-adhering tape at $t=0$, and the critical values of $\sigma_{0}$ and
$\theta_{0}$ for peeling initiation are given by Kendall's equation
\cite{Kendall1971,Menga2018,Menga2020}%
\begin{equation}
\frac{\sigma_{0}^{2}}{2E_{\infty}}+\sigma_{0}\left(  1-\cos\theta_{0}\right)
-\frac{\gamma}{d}=0{,}
\label{Starting solution 1}
\end{equation}
and, from Eq. (\ref{geometric}) with $s_{d}=0$
\begin{equation}
\frac{\cos\phi-\cos\theta_{0}}{\cos\theta_{0}}=\frac{\sigma_{0}}{E_{\infty}}{.}
\label{Starting solution 2}%
\end{equation}

Finally, the critical force is calculated as $P_{0}=A_{t}\ \sigma_{0}\ \sin\theta_{0}$. For $t>0$, the peeling process evolution follows Eqs. (\ref{geometric},\ref{Energy Balance}) and is calculated by exploiting the numerical procedure outlined in Appendix \ref{numerical implementation}.

\begin{figure}[ptbh]
\begin{center}
\includegraphics[width=1\textwidth]{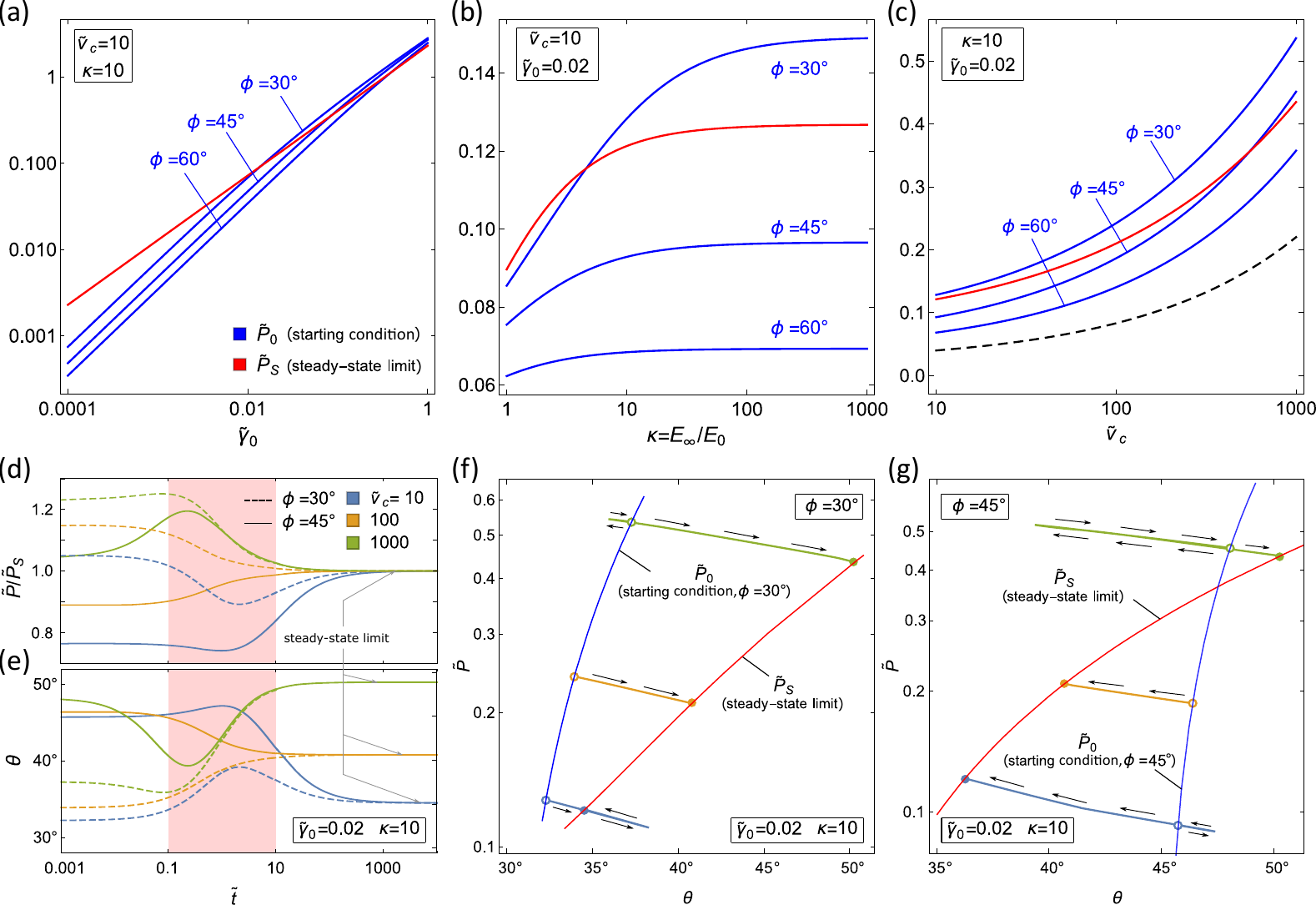}
\end{center}
\caption{Top row: the initial $\tilde{P}_{0}$ and long-term $\tilde{P}_{S}$ dimensionless peeling force as functions of the dimensionless adhesion energy $\tilde{\gamma}_{0}$ (a), the viscoelastic parameter $\kappa=E_{\infty}/E_{0}$ (b), and the dimensionless peeling front velocity $\tilde{v}_{c}$ (c). In figure (c), the rate-dependent dimensionless adhesion energy $\tilde{\gamma}$ is also shown for comparison (dashed line). Notably, $\kappa=1$ corresponds to elastic tapes. Bottom row: the time-history of the normalized peeling force $\tilde{P}/\tilde{P}_{\mathrm{S}}$ (d) and peeling angle $\theta$ (e) for different values of the dimensionless peeling front velocity $\tilde{v}_{c}$ and initial undeformed configurations. Non-monotonic behavior occurs in the red regions, i.e. for $t\approx\tau$. Transient diagrams $\tilde{P}$ versus $\theta$ are shown for different values of $\tilde{v}_{c}$ and for $\phi=30\degree$ (f) and $\phi=45\degree$ (g). Blue and red curves represent the starting condition and the steady-state limit, respectively. Black arrows indicate the time evolution of the process.}  
\label{fig5}%
\end{figure}

Comparing the critical peeling force $P_0$ with the long-term limit $P_S$ allows to differentiate from toughening and weakening overall peeling behaviors. This is done in the top row of Fig. \ref{fig5}, where we consider the effect of (a) the dimensionless adhesion energy $\tilde{\gamma}_{0}$, (b)
the viscoelastic parameter $\kappa=E_{\infty}/E_{0}$, and (c) the
dimensionless peeling front velocity $\tilde{v}_{c}$.
While in the elastic V-peeling case, the toughest behavior always occurs in the steady-state limit, as clearly shown for $\kappa=1$ in Fig. \ref{fig5}b where $P_{\mathrm{S}}>P_{\mathrm{0}}$, the same principle cannot be generalized to the viscoelastic case, where the value of both $P_0$ and $P_S$ depends on the tape relaxation process and, as a consequence, on the specific combination of the parameters $\tilde{\gamma}_{0}$, $\kappa\,$, and $\tilde{v}_{c}$, as shown in Figs. \ref{fig5}a,b,c. 
Moreover, according to Eqs. (\ref{Starting solution 1}-\ref{Starting solution 2}), the critical starting force $P_0$ depends on the undeformed tape angle $\phi$, specifically leading to tougher peeling initiation with $\phi$ reducing. 
In agreement with theoretical \cite{Kendall1971} and experimental results \cite{Xia2013}, stiffer tapes entail higher peeling forces, as shown in Fig. \ref{fig5}b; nonetheless, at very large values of $\kappa$ both $P_0$ and $P_S$ are almost constant and the Rivlin \cite{Rivlin1944} solution for rigid tapes is asymptotically approached.

The bottom row of Fig. \ref{fig5} shows the peeling transient evolution from start to steady-state behavior. The most important feature is that V peeling of viscoelastic tapes may present non-monotonic trends of both $\theta$ and $\tilde{P}$, in contrast with results achieved for the elastic case in Refs. \cite{Afferrante2013,Menga2018,Menga2020}. More in detail, focusing on \ref{fig5}e, given the undeformed tape angle $\phi$, the tape angle at peeling start $\theta_0$ increases with $\tilde{v}_{c}$, as expected from Eq. (\ref{Starting solution 2}) and Fig. \ref{fig5}c showing $\theta_{0}$ increasing with $\tilde{P}_{0}$ and $\tilde{P}_{0}$ increasing with $\tilde{v}_{c}$, respectively. 
Following Eq. (\ref{geometric}), once the peeling starts, the value of $\theta(t)$ depends on the interplay between (i) the peeling front propagation, causing a linear increase of $s_{d}$, and (ii) the detached tape relaxation, increasing the term $\Delta L$. For $\tilde{t}\ll1$, a rough estimation of $\dot{\theta}$ can be derived from Eq. (\ref{geometric}) as $\dot{\theta}\propto d\left(  \Delta L\right)  /dt-\beta
v_{c}$, with $\beta=\beta\left(  \theta_{0}\right)  $ being a monotonically increasing function; indeed, in agreement with Fig. \ref{fig5}e, f, and g, the peeling angle $\theta$ at $\tilde{t}\ll1$ can either decrease (at high velocity, i.e. $\tilde{v}_{c}\approx1000$) or increase (at low velocity, i.e. $\tilde{v}_{c}\approx10$), while in the long-term limit $\theta(\tilde{t}\gg1)\approx\theta_{\mathrm{S}}$ eventually leading to non-monotonic behavior, depending on the specific value of $\theta_{\mathrm{S}}$.
The normalized peeling force $\tilde{P}%
/\tilde{P}_{S}$ in Fig. \ref{fig5}d, f, and g is non-monotonic, as well, since high values of $\theta$ lead to low values of $\tilde{P}/\tilde{P}_{S}$ and vice versa, as expected \cite{Kendall1971,Afferrante2013,Ceglie2022}.

\subsection{Constant peeling force}

In this section, we investigate the viscoelastic V peeling behavior
under a constant peeling force $P$, such as under the action of a dead weight. Surprisingly, the results show that the peeling process can either start and indefinitely propagate, start and then stop after some time, or not even start at all, depending on the value of $P$ and initial tape geometry (i.e., the undeformed angle $\phi$ and length $L_i$). The boundaries between these qualitatively different behaviors depend on the peeling front velocity, which is not known \textit{a priori} in this case; therefore, the critical (minimum) forces for peeling start and steady-state propagation must be sought for both $v_c\gg d/\tau$ and $v_c\ll d/\tau$ assuming $\gamma\approx\gamma_0$, as for real thin tapes $\tilde{v}_{\gamma}\approx 10 d/\tau$. As discussed at the end of Section \ref{Formulation}, in the former case, critical loads for peeling start $P_1$ and steady-state propagation $P_2$ are given by Eqs. (\ref{Starting solution 1},\ref{Starting solution 2}) and Eqs. (\ref{steady state 1},\ref{steady state 2}), respectively, with $\gamma\approx\gamma_{0}$. On the contrary, in the latter case (i.e., for $v_c\ll d/\tau$), the critical forces $P_3$ (start) and $P_4$ (steady-state propagation) are given by the same equations with the high-frequency modulus $E_{\infty}$ replaced by the low-frequency one $E_{0}$ and, again, with $\gamma\approx\gamma_{0}$. 

\begin{figure}[ptbh]
\begin{center}
\includegraphics[width=1\textwidth]{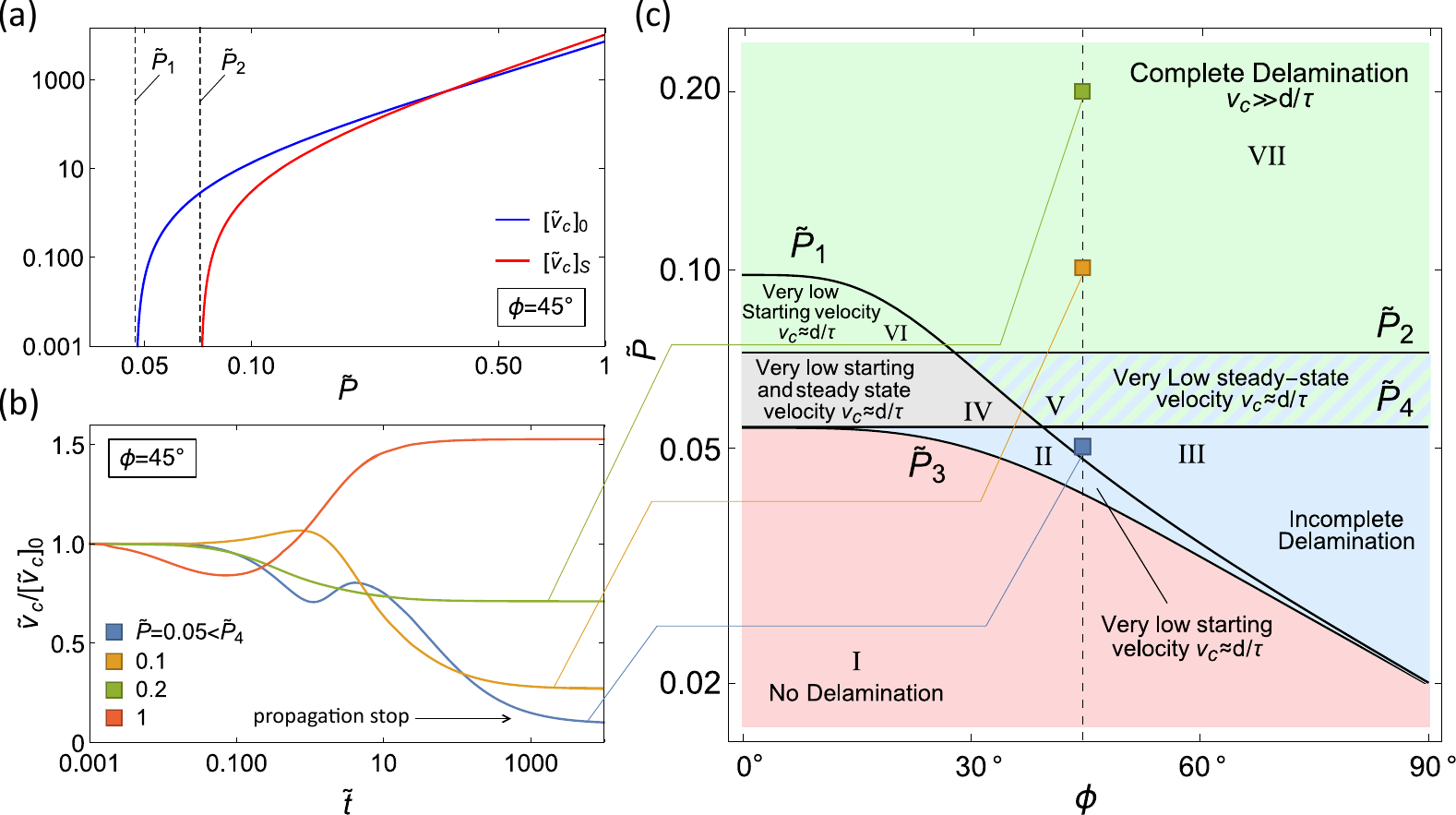}
\end{center}
\caption{(a) The initial $[\tilde{v}_c]_{0}$ and long-term $[\tilde{v}_c]_{\mathrm{S}}$
dimensionless peeling front velocity as functions of the dimensionless peeling force
$\tilde{P}$. (b) The time-history of the normalized peeling front velocity $\tilde{v}_c/[\tilde{v}_c]_{0}$ for different values of dimensionless peeling force $\tilde{P}$. (c) The state map of the possible peeling behavior as a function of the dimensionless applied peeling force $\tilde{P}$ and undeformed tape angle $\phi$. 
$\tilde{P}_{1}$ and $\tilde{P}_{2}$ are the critical (minimum) force for peeling start and steady-state propagation calculated with $v_c\gg d/\tau$, while $\tilde{P}_{3}$ and $\tilde{P}_{4}$ refer to the same critical conditions when $v_c\ll d/\tau$ (further know are given in the text). Results refer to $\kappa=10$ and $\tilde{\gamma}_0=0.02$.}%
\label{fig7}%
\end{figure}

The map in Fig. \ref{fig7}c shows the possible peeling behaviors as discussed above as functions of the dimensionless applied peeling force $\tilde{P}$ and undeformed tape angle $\phi$ (we assume $L_i=100d$ for all calculations). 
Specifically, in region I the peeling does not propagate, in region II-III the peeling propagation starts and then stops after some time, and in regions IV-V-VI-VII the peeling propagates indefinitely approaching the steady-state regime (though in IV and V the long-term velocity is lower than $d/\tau$). 
The present model predictions are rigorously valid in regions I and VII; nonetheless, qualitative insight can also be inferred for regions III and V, as a steady-state propagation with very low velocity (about $d/\tau\approx10^{-4}$ m/s) qualitatively corresponds to a peeling stop (i.e., as in case III), for region VI, where $v_{c}\approx d/\tau$ when peeling starts and then rapidly increases, and for region II, as the peeling cannot be sustained in steady-state conditions regardless of the starting behavior. 
Finally, region IV cannot be accounted for in the present framework, as the specific viscoelastic behavior close to the peeling front does really matter throughout the whole process evolution. In most cases, real systems belong to the first scenario, with $v_{c}\gg d/\tau.$

The peeling front velocity calculated at the process start $[v_c]_0$ and in the long-term steady-state $[v_c]_S$ limit are shown in Fig. \ref{fig7}a as functions of $\tilde{P}$. 
More interestingly, two different peeling behaviors are shown in Fig. \ref{fig7}b, with respect to the time-history of the normalized peeling front velocity $\tilde{v}_{c}/\left[  \tilde{v}_{c}\right]
_{0}$ for different values of $\tilde{P}$ belonging to regions VII and III. 
In the latter case (blue curve), for $\tilde{P}< \tilde{P_4}$, the peeling front velocity decreases down to full stop. In the other cases, all belonging to region VII, after the initial non-monotonic behavior, in the long-term a steady-state behavior is approached, with endless propagation occurring at velocity $[v_c]_S$.

\subsection{Constant pulling velocity}

The final case we deal with is with the tape being pulled at a constant velocity $\tilde{v}_{P}$. In this case, the start of peeling propagation does not coincide with the application of the pulling velocity. Indeed, before peeling starts, the stress $\sigma$ in the non-adhering tape must increase from zero to a certain critical value $\sigma_{cr}$. The energy-based procedure to calculate such a critical condition is given in Appendix \ref{Appendix B}.

\begin{figure}[ptbh]
\begin{center}
\includegraphics[width=1\textwidth]{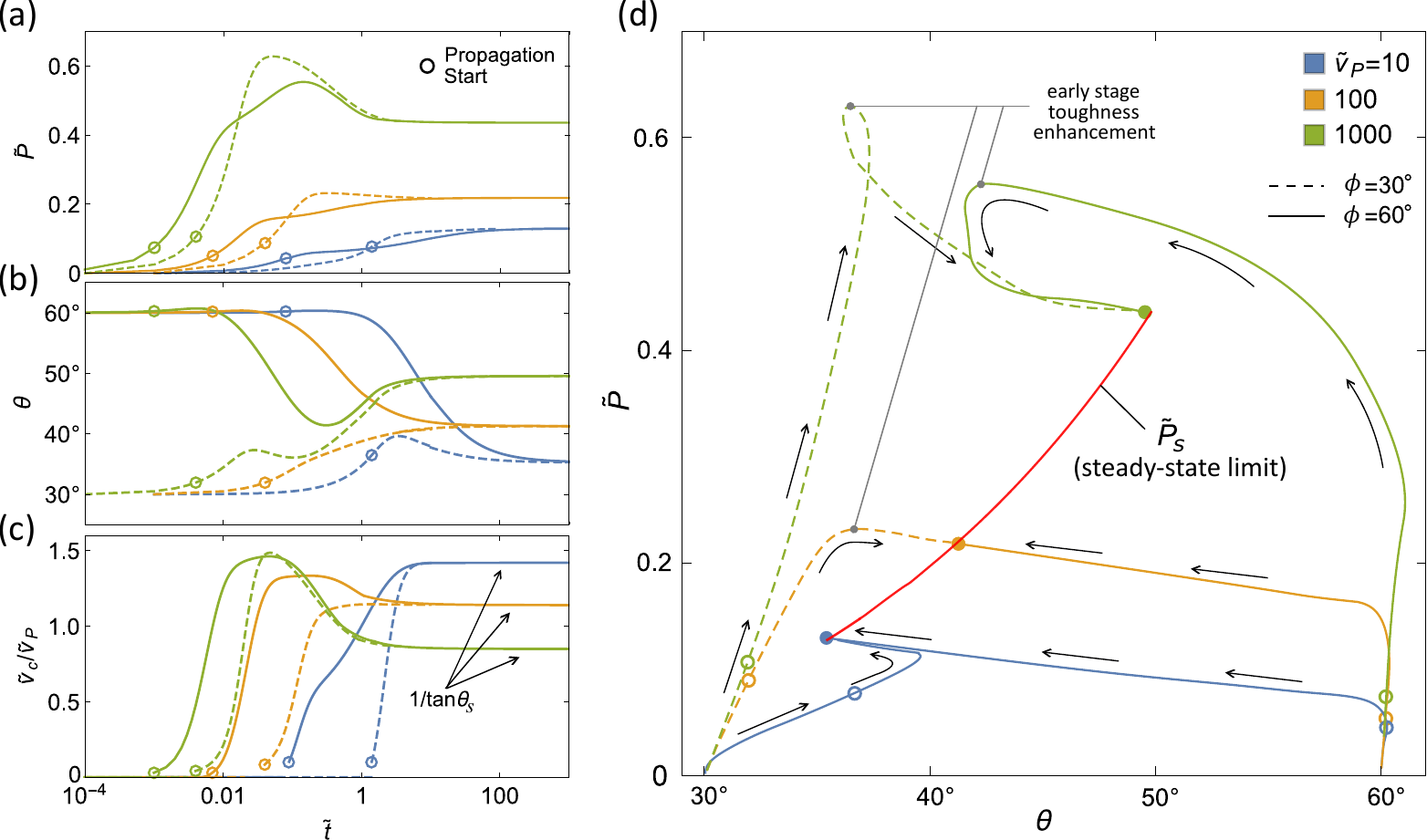}
\end{center}
\caption{The time-history of the dimensionless peeling force
$\tilde{P}$ (a), the peeling angle $\theta$ (c), and the peeling velocity ratio $\tilde{v}_{c}/\tilde{v}_{P}$ (c), together with the transient diagram $\tilde{P}$  vs. $\theta$ (d) for different values of the dimensionless pulling velocity $\tilde{v}_{P}$. Two undeformed tape angles are considered. The circles indicate the instant when peeling front propagation starts. The red curve in (d) represents the steady-state peeling limit, and black arrows indicate the time evolution of the process. Results refer to $\kappa=10$ and $\tilde{\gamma}_0=0.02$.}%
\label{fig10}%
\end{figure}

In Figs. \ref{fig10}a, b, and c, we show, respectively, the time-history of $\tilde{P}$, $\theta$ and
$\tilde{v}_{c}/\tilde{v}_{P}$, for different values of the dimensionless
pulling velocity $\tilde{v}_{P}$ and undeformed angle $\phi$. Circles indicate the peeling start, which increases with $\phi$ reducing, as expected from Eq. (\ref{sigma critic}). In the long-term limit, steady-state propagation occurs, with $\left[
v_{c}\right]  _{\mathrm{S}}/\tilde{v}_{P}=1/\tan\theta_{\mathrm{S}}$ according to Eq. (\ref{steady state vel}). The most interesting result from Fig. \ref{fig10}a, is that the peeling force $\tilde{P}$ may present a maximum at the early stage, right after the peeling start, at a relatively high pulling velocity $\tilde{v}_{P}$. This is also shown in Fig. \ref{fig10}d, clearly indicating that this is associated with a temporary reduction of the peeling angle $\theta$ which results from a fast increase of $v_c$ before viscoelastic relaxation occurs (i.e., for $\tilde{t}\ll 1$).
Such a peculiar feature may partially link to the superior adhesive performance of V-shaped natural systems, such as spider webs \cite{Cranford2012} and mussels byssus \cite{Qin2013}, under the action of high-speed (impact) loading conditions. In the latter case, for instance, Cohen et Al. \cite{Cohen2019} have shown that the single byssus is highly stretchable, due to the heterogeneous filament structure (a system of nonlinear swollen springs); here, we suggest that also the interplay between byssus rheology and V-shaped multiple threads geometry (see Fig. \ref{image}c) may contribute to the observed tougher adhesive response under dynamic loads \cite{Qin2013}.

\section{Conclusions}

In this study, we model the peeling behavior of a viscoelastic thin tape arranged in V-shaped peeling configuration. Specifically, the velocity-dependent condition for peeling front propagation is found in terms of energy balance between the work per unit time done by the internal stress in the tape, the external forces acting on the system, and the surface adhesion forces. An \textit{ad hoc} numerical procedure is derived to predict the time-evolution of the peeling process, taking into account the time-varying viscoelastic relaxation of the detached tape. We consider three possible physical scenarios for peeling propagation: constant peeling front velocity, constant peeling force, and constant pulling velocity at the tape tip.

In the long-term limit, the peeling propagation asymptotically approaches a steady-state elastic-like behavior, regardless of the specific controlled parameter. However, the initial transient peeling behavior is strongly affected by the tape viscoelasticity and undeformed geometry, and presents non-monotonic time evolution of the peeling force and angle. More in detail, when a constant force is applied, we found that the peeling can either endlessly propagate, start and stop after some time, or not even start. Which of these scenarios occurs seems to depend only on the applied force value and undeformed non-adhering tape geometry (angle and length).

More surprisingly, when the pulling velocity at the tape tip is assigned, as in the case of impact loads acting on an attached object, the peeling propagation is delayed against the instant of force application, and the force required to sustain the peeling propagation (i.e., the peeling toughness) can be temporarily larger than in stationary conditions. This mechanism might be qualitatively related to the high-speed superior adhesive performance observed in several natural systems.

\begin{acknowledgments}
This work was partly supported by the Italian Ministry of Education,
University and Research under the Programme \textquotedblleft Department of
Excellence\textquotedblright\ Legge 232/2016\textquotedblright\ and partly by
the European Union - NextGenerationEU (National Sustainable Mobility Center
CN00000023, Italian Ministry of University and Research Decree n. 1033 -
17/06/2022, Spoke 11 - Innovative Materials \& Lightweighting). The opinions
expressed are those of the authors only and should not be considered as
representative of the European Union or the European Commission's official
position. Neither the European Union nor the European Commission can be held
responsible for them.
\end{acknowledgments}

\appendix
\counterwithin{figure}{section}
\section{Numerical calculation of peeling process evolution}

\label{numerical implementation}

\begin{figure}[ptbh]
\begin{center}
\includegraphics[width=0.6\textwidth]{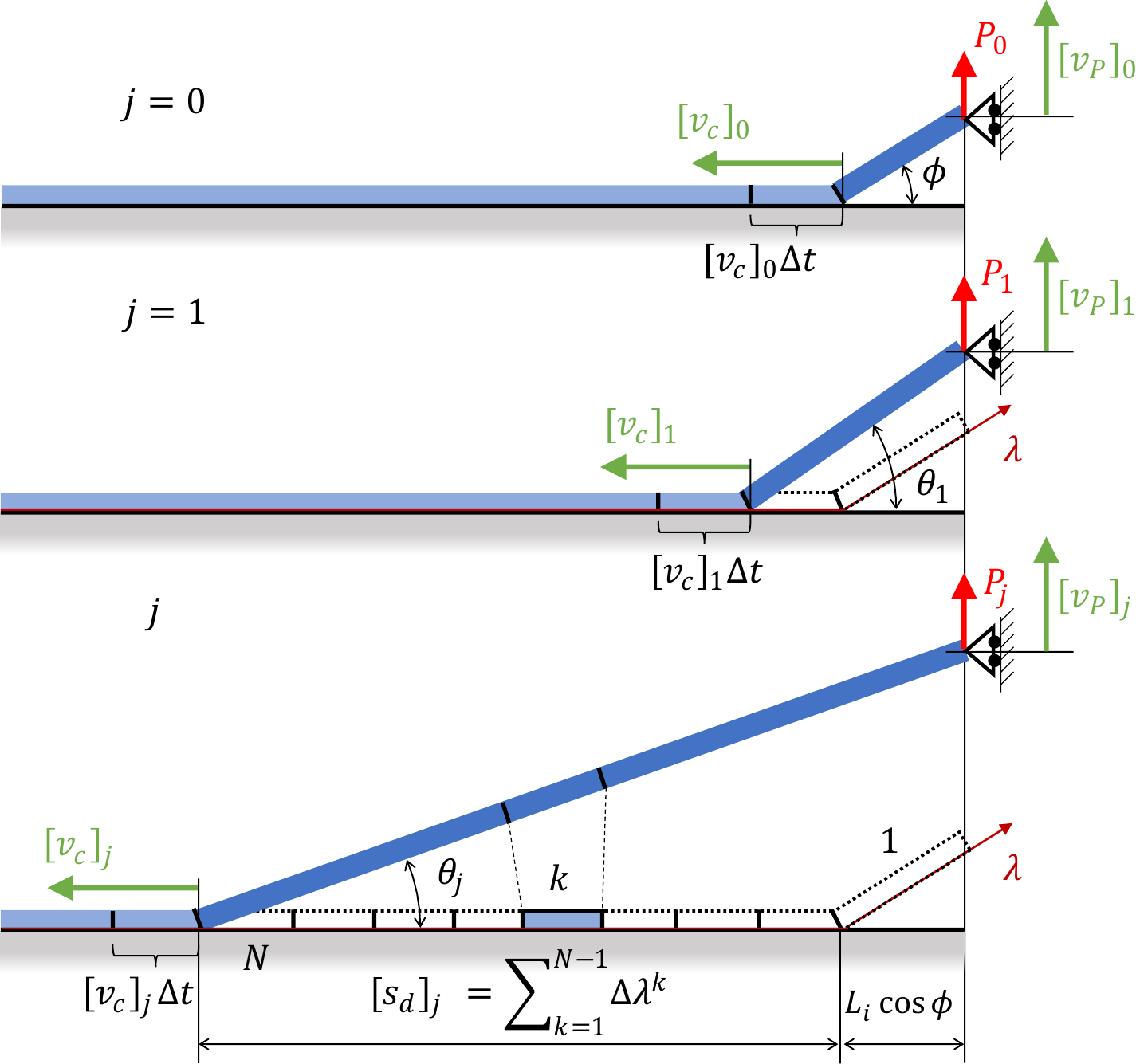}
\end{center}
\caption{Scheme of the tape discretization: at the generic $j$-th time instant, the tape mesh is updated by including a freshly detached element of undeformed length $\Delta\lambda=\left[  v_{c}\right]  _{j}\Delta t$.}
\label{fig2}
\end{figure}

The process evolution can be calculated by numerical integration of Eqs.
(\ref{Energy Balance}, \ref{geometric}). However, dealing with a viscoelastic
tape, the governing equations depend on the entire time-history of the process
and the solution of Eqs. (\ref{Energy Balance}, \ref{geometric}) must be
iteratively sought by successively updating the system configurations as the peeling front moves.

The numerical method is based on uniform time discretization, with time step
$\Delta t\ll\tau$, and non-uniform tape mesh. In the following discussion, the
notation $\eta_{j}^{k}$ represents the discrete value of the generic quantity
$\eta$ at the $j$-th time instant in the $k$-th tape element. Only the
non-adhering tape is discretized, and the mesh is updated at each time step so
that, at the generic time $t_{j}=j\Delta t$, an element of (undeformed) length
$\Delta\lambda=\left[  v_{c}\right]  _{j}\Delta t$ is added to the mesh due to peeling front motion. The resulting non-uniform "incremental" mesh has two primary
advantages: (i) the total number of elements does not need to be fixed a
priori; (ii) only the non-adhering tape is discretized, and the computational
cost is reduced. More in detail, referring to Fig. (\ref{fig2}), $N_{j}=j+1$
is the total number of tape elements at the $j$-th time instant, with the
first element $\Delta\lambda^{1}$ being the initial non-adhering tape (i.e.,
$\Delta\lambda^{1}=L_{i}$) and the $N$-th element $\Delta\lambda^{N_{j}}$
being last detached element ($\Delta\lambda^{N_{j}}=\Delta t\left[
v_{c}\right]  _{j-1}$). Therefore, the detached tape projection $s_{j}$ is
given by
\begin{equation}
s_{j}=L_{i}\cos\phi+\left[  s_{d}\right]  _{j}=L_{i}\cos\phi+\sum_{k=2}%
^{N_{j}}\Delta\lambda^{k}{.}
\end{equation}

The discrete form of Eqs. (\ref{wad},\ref{wf}) is
\begin{align}
\left[  W_{ad}\right]  _{j}  &  =-\left[  v_{c}\gamma\right]  _{j}w
\label{wad_disc}\\
\left[W_{P}\right]  _{j}  &  =A_{t}\ \,\left[  v_{P}\sigma\right]  _{j}%
\sin\theta_{j} \label{wp_disc}%
\end{align}
where, from Eq. (\ref{velocity}),
\begin{equation}
\left[  v_{P}\right]  _{j}=\left[  v_{c}\right]  _{j}\,\tan\theta_{j}%
+\frac{s_{j}}{\cos^{2}\theta_{j}}\frac{\theta_{j}-\theta_{j-1}}{\Delta t}{.}
\end{equation}
Integrating by parts, Eq. (\ref{constitutive}) is rewritten as
\begin{equation}
\varepsilon(\lambda,t)=\frac{\sigma(\lambda,t)}{E_{\infty}}+\int
\nolimits_{-\infty}^{t}\mathcal{\dot{J}}(t-t^{\prime})\sigma(\lambda
,t^{\prime})dt^{\prime}{,}
\label{constitutive2}
\end{equation}
where we used $\mathcal{J}(0)=E_{\infty}^{-1}$, and $\sigma(\lambda
,-\infty)=0$; the discrete form of Eq. (\ref{constitutive2}) gives the
elongation of the generic $k$-th element of the non-adhering tape at the
$j$-th time instant as
\begin{equation}
\varepsilon_{j}^{k}=\frac{\sigma_{j}^{k}}{E_{\infty}}+\Delta t\sum_{h=0}^{j}\mathcal{\dot{J}}_{j-h}\sigma_{h}{,}
\end{equation}
with $\sigma_{j}^{k}=\sigma_{j}=P_{j}/\left(  A_{t}\sin\theta_{j}\right)  $ for all the tape elements.

To calculate $W_{in}$, we observe that the term $\partial\varepsilon/\partial
t(\lambda,t)\,$\ diverges at the peeling front (i.e., for $\lambda
\rightarrow\lambda_{c}\left(  t\right)  $), as we assume a step-change in the
tape stress $\sigma$. In this case, the discretized form of Eq. (\ref{win}) can
be conveniently rewritten as
\begin{equation}
\frac{\left[  W_{in}\right]  _{j}}{A_{t}}=-\sigma_{j}\dot{\varepsilon}_{j}%
^{N}\Delta\lambda^{N}-\sigma_{j}\sum_{k=1}^{N-1}\dot{\varepsilon}_{j}
^{k}\Delta\lambda^{k}{,}
\label{win_separato}
\end{equation}
where, according to \cite{Ceglie2022,Persson2021}, since $v_{c}\gg d/\tau$,
the first right-hand side term can be calculated as
\begin{equation}
\sigma_{j}\dot{\varepsilon}_{j}^{N}\Delta\lambda^{N}=\frac{\left[  \sigma
^{2}v_{c}\right]  _{j}}{2E_{\infty}}{.} 
\label{win_first}%
\end{equation}
Combining Eqs. (\ref{win_separato}, \ref{win_first}), the discretized form of
Eq. (\ref{win}) is%
\begin{equation}
\frac{\left[  W_{in}\right]  _{j}}{A_{t}}=-\frac{\left[  v_{c}\sigma
^{2}\right]  _{j}}{2E_{\infty}}-\sigma_{j}\sum_{k=1}^{N-1}\left[
\frac{\varepsilon_{j}-\varepsilon_{j-1}}{\Delta t}\Delta\lambda\right]  ^{k}
\label{win_disc}%
\end{equation}

Finally, using Eqs. (\ref{wad_disc},\ref{wp_disc},\ref{win_disc}) in Eq.
(\ref{Energy Balance}) gives the discrete form for the instantaneous energy
balance equation
\begin{equation}
\left[  W_{P}\right]  _{j}+\left[  W_{in}\right]  _{j}+\left[  W_{ad}\right]
_{j}=0 \label{energy_balance_disc}%
\end{equation}
and, from Eq. (\ref{geometric}) we have
\begin{equation}
\frac{s_{j}}{\cos\theta_{j}}=\sum_{k=1}^{N}\left(  1+\varepsilon_{j}%
^{k}\right)  \Delta\lambda^{k}{,}
\label{geometric discrete}
\end{equation}
where $\left(  1+\varepsilon_{j}^{k}\right)  \Delta\lambda^{k}$ is the deformed length of the generic $k$-th element at the $j$-th time instant. The process time evolution is obtained from an iterative algorithm based on the Newton-Rapshon method that solves Eqs. (\ref{energy_balance_disc},\ref{geometric discrete}) for the unknown peeling quantities at each time instant.

\section{Starting condition under constant pulling velocity}
\label{Appendix B}

\begin{figure}[ptbh]
\begin{center}
\includegraphics[width=1\textwidth]{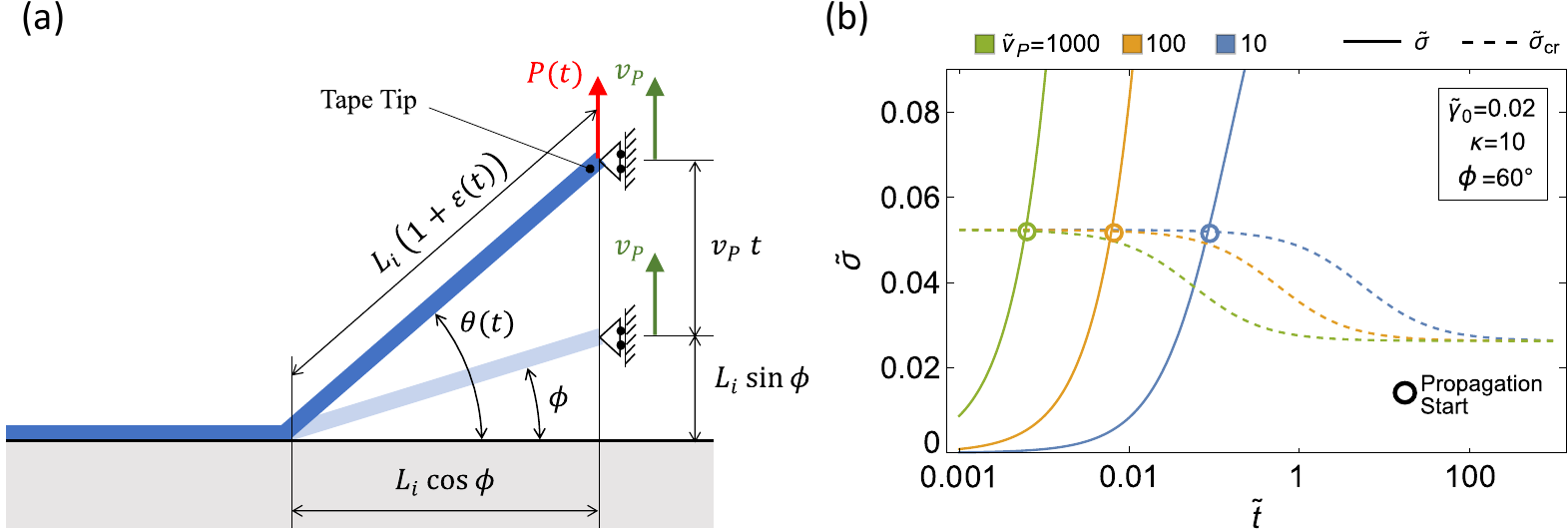}
\end{center}
\caption{(a) A schematic of the tape in undeformed condition,
and at a generic time $t<t^{\ast}$, i.e. before the peeling propagation starts. (b)
The dimensionless stress $\tilde{\sigma}$ in the tape (solid curves) and the dimensionless critical stress $\tilde{\sigma}_{cr}$ required to start the peeling propagation (dashed
curves) as functions of the dimensionless time $\tilde{t}$ for different
dimensionless tape tip velocity $\tilde{v}_{P}$. Circles indicate the
instant of propagation start.}%
\label{fig9}%
\end{figure}

We consider that at time $t=0$ the tape tip is pulled at a constant velocity $v_{P}$. In this case, the deformation $\varepsilon(t)$ and stress $\sigma(t)$ in the non-adhering tape monotonically increase, and peeling front propagation starts at time $t^{\ast}$ when $\sigma(t^{\ast})=\sigma_{cr}(t^{\ast})$. The value $\sigma_{cr}$ depends on the energy balance
\begin{equation}
\frac{\sigma_{cr}^{2}}{2E_{\infty}}+\sigma_{cr}[1-\cos\theta]=\frac{\gamma_{0}}{d}{,}
\label{sigma critic}
\end{equation}
where we assumed that $v_c(t^{\ast})\ll v_\gamma$. 

According to Fig. \ref{fig9}a, before peeling front propagation (i.e., for $t<t^{\ast}$), the peeling angle is given by
\begin{equation}
\tan\theta=\frac{L_{i}\sin\phi+v_{P}t}{L_{i}\cos\phi}{,}
\label{theta t}
\end{equation}
where $v_{P}t$ is the tape tip vertical displacement at the generic time $t$. Similarly, since $L_{i}+\Delta L=L_{i}\cos\phi/\cos\theta$ is the deformed tape length, the tape uniform deformation before peeling propagation is
\begin{equation}
\varepsilon(t)=\frac{L_{i}+\Delta L}{L_{i}}-1=\frac{\cos\phi}{\cos\theta}-1{.}
\end{equation}
Finally, using the viscoelastic constitutive equation, we can calculate the uniform stress in the detached tape as
\begin{equation}
\sigma(t)=\int_{-\infty}^{t}\mathcal{R}(t-t^{\prime})\dot{\varepsilon}(t^{\prime})dt^{\prime}{,}
\label{sigma t}
\end{equation}
where $\mathcal{R}$ is the stress-relaxation function given by
\begin{equation}
\mathcal{R}(t)=E_{0}+(E_{\infty}-E_{0})e^{t/\tau_{r}}{,}
\end{equation}
with $\tau_{r}=\tau/(1+\Delta)$ being the relaxation time, and $\Delta
=E_{\infty}/E_{0}-1$ being the relaxation strength.

The peeling starting condition $\sigma(t^{\ast})=\sigma_{cr}(t^{\ast})$ is then obtained by simultaneously solving Eqs. (\ref{sigma critic}-\ref{sigma t}). In Figure \ref{fig9}b we show the time-history of $\tilde{\sigma}(t)$ and $\tilde{\sigma}_{cr}(t)$, for different dimensionless pulling velocities $\tilde{v}_{P}$. Increasing $\tilde{v}_{P}$ leads to faster stress increase in the viscoelastic tape, thus peeling propagation starts sooner.

Once the peeling front propagation starts, the numerical algorithm described in Appendix \ref{numerical implementation} can be used to calculate the time evolution of the peeling process, with $t^{\ast}$ corresponding to $j=0$. In the present formalism, the peeling front velocity $v_{c}(t^{\ast})$ at the instant of the peeling propagation start cannot be exactly determined; however, in the reasonable assumption for practical applications that $v_{P}/\tan\theta_{\mathrm{S}}=[v_{c}]  _{\mathrm{S}}\gg d/\tau$, the system rapidly approaches the conditions $v_{c}>d/\tau$ at the very early stage of peeling propagation. As a consequence, we set $\tilde{v}%
_{c}\left(  t^{\ast}\right)  \approx1$.

\end{document}